\documentclass[twocolumn,prl,showpacs]{revtex4}

\usepackage{graphicx}
\usepackage{rotating}
\usepackage{amsmath}
\usepackage{amsfonts}
\usepackage{amssymb}
\usepackage{enumerate}
\usepackage{longtable}
\usepackage{dcolumn}
\usepackage{bm}

\begin{document}

\newcommand{\be}{\begin{equation}}
\newcommand{\ee}{\end{equation}}
\newcommand{\bn}{\begin{eqnarray}}
\newcommand{\en}{\end{eqnarray}}

\title{Theory of Magnetic Fluctuations in Iron Pnictides}

\author{L. Craco and M.S. Laad}
\affiliation{Max-Planck-Institut f\"ur Physik komplexer Systeme,
01187 Dresden, Germany}

\date{\rm\today}

\begin{abstract}
Magnetic fluctuations in an unconventional superconductor (U-SC)  
can distinguish between distinct proposals for the symmetry of the 
order parameter.  Motivated thereby, we undertake a study magnetic 
fluctuations in Iron pnictides, tracking their evolution from the 
incoherent normal, pseudogapped metal, to the U-SC state. Within 
our proposal of extended-$s$-plus $s_{xy}$ inplane gap with 
proximity-induced out-of-plane line nodes, (i) we describe the 
evolution of the spin-lattice relaxation rate, from a non-Korringa 
form in the normal state, to a power-law form in the U-SC in 
good agreement with experiment, and (ii) we predict a sharp 
resonance in the U-SC state along $(\pi,\pi)$, but not along 
$(\pi/2,0)$, along with modulated $c$-axis intensity in inelastic 
neutron scattering work as a specific and testable 
manifestation of our proposal.             
\end{abstract}
    

\maketitle

The precise mechanism of unconventional superconductivity (U-SC) in the 
recently discovered Iron Pnictides (FePn) is presently a hotly debated 
issue~\cite{[1]}. While many physical responses are reminiscent of 
cuprates~\cite{[2]}, FePn are metals, albeit presumably proximate to 
a Mott insulator. Moreover, relevance of {\it all} $d$ orbitals in 
FePn considerably complicates determination of the pair symmetry.

Study of magnetic fluctuations in an U-SC can help unearth the symmetry 
of the SC order parameter, as shown by detailed studies for cuprates~\cite{[3]}.
In the FePn, NMR studies already reveal normal state pseudogap 
behavior~\cite{[4]} and U-SC. The spin-lattice relaxation rate, $T_{1}^{-1}$,
shows marked deviation from the linear-in-$T$ Korringa form expected from a 
Fermi liquid and smoothly {\it decreases} for $T<2T_{c}$, where $T_{c}$ is the
SC transition temperature.  At very low $T$,
one finds $T_{1}^{-1}\simeq T^{n}$ with $n\ne 3,5$, indicating line nodes in 
the SC gap.  However, other probes reveal anisotropic, albeit fully 
gapped, structure of the in-plane gap function.  Thus, extant data imply that,
either one has out-of-plane line nodes, as in $Sr_{2}RuO_{4}$~\cite{[5]}, or
disorder effects in a $s_{\pm}$-SC produce the observed behavior~\cite{[6]}.
In the extended-$s$ wave idea, disorder is argued to lift the nodal structure,
again giving similar behavior~\cite{[7]}.  The issue is thus controversial: 
while ARPES data are inconclusive regarding existence of nodes on
electron-like FS sheets~\cite{[8]}, a penetration depth study, at least in 
$SmO_{1-x}FeAsF_{x}$, shows {\it smooth} angular variation of the in-plane 
gap~\cite{[9]}.
While inelastic neutron scattering (INS) work does reveal a low-energy 
resonance structure in the U-SC state 
for ${\bf Q}=(\pi,\pi)$~\cite{[10]}, more detailed map of the INS response
in ${\bf q}$-space awaits future work.  To date, we are aware of one study 
where the dynamical spin susceptibility, $\chi''({\bf q},\omega)$, has been 
measured for the 122 FePn, showing that the SC gap has in-plane smooth angular 
variation {\it and} an out-of-plane cos$(k_{z}c)$ 
component~\cite{[11]}.  

Extant theoretical works have studied these issues using effective model
Hamiltonians, both in the weak~\cite{[12]} and strong~\cite{[13],[14]} coupling
limits.  In the itinerant approach, the magnetic fluctuations have been
computed within HF-RPA.  For $s_{\pm}$ pairing, a sharp resonance for 
${\bf Q}=(\pi,\pi)$ in INS is predicted below $T_{c}$~\cite{[12]}, while 
no such feature is arises for $s$, ex-$s$, and $d$ wave-pairing, or for 
${\bf q}\ne {\bf Q}$. To get the power-law-in-$T$ behavior in NMR and 
$\mu$SR in the $s_{\pm}$ idea, it is necessary to consider (strong) 
disorder effects in a two- or four-band model.  Again, the situation is 
controversial.  For $LaFePO$, the penetration depth, 
$\lambda(T)\simeq T^{1.2}$~\cite{[15]}, 
while a similar study on a wide range of samples of different FePn
found a seeming universality in $\lambda(T)$; this mitigates against  the 
disorder effects~\cite{[16]}.  {\it If} this is true, one must consider both
the NMR and INS data within a theoretical scenario with out-of-plane line 
nodes in the SC gap, since in-plane nodes seem to be ruled out by extant
tunnelling data~\cite{[17]}.

Recently, based on inputs from the correlated normal state electronic 
structure and rigorous symmetry arguments, we proposed a specific gap 
function with sizable in-plane angular variation (but no nodes) {\it and} 
inter-band proximity induced out-of-plane line nodes~\cite{[18]}. In 
contrast to the itinerant picture, our proposal is based on a strong 
correlation view of FePn.  Here, we investigate the NMR and INS response 
within such a correlated approach, using the {\it full}, multiband 
spectral functions for {\it all} $d$ orbitals. LDA+DMFT can readily 
access the intermediate coupling regime relevant for FePn~\cite{[19]}.  
We show how our proposal gives a quantitative account of the NMR 
$T_{1}^{-1}$ over the whole $T$ range, and makes specific predictions 
with regard to the observation of the low-energy {\it dispersive} 
resonance in the INS intensity below $T_{c}$.         

The central quantity of interest is the dynamical spin susceptibility, 
$\chi({\bf q},\omega)=\sum_{a,b}\chi_{ab}({\bf q},\omega)$, where $a,b$ 
are {\it all} $d$-orbital indices, and ${\bf q},\omega$ are the momentum and 
energy transfers in INS.  Viewing FePn as strongly correlated 
systems with $U=4.0$~eV, $U'=2.6$~eV and $J_{H}=0.7$~eV, we construct 
$\chi({\bf q},\omega)$ in terms of the {\it full} 
LDA+DMFT propagators computed in earlier work~\cite{[18],[20]}.  Very good 
quantitative agreement between LDA+DMFT and key experiments in {\it both}, the
normal and U-SC states, has been shown there, lending strong support for our
choice.  The prescription is simple: replace the band Green functions used
in weak-coupling approaches~\cite{[12]} by their LDA+DMFT counterparts.
This ensures that the dynamical aspect of strong, local, multi-orbital (MO) 
correlations is included from the outset.  

For a MO-system, after replacing the {\it bare} $G_{aa}({\bf k},\omega)$ with 
$G_{aa}({\bf k},\omega)\equiv G_{aa}^{LDA+DMFT}({\bf k},\omega)=
[\omega-\epsilon_{ka}-\Sigma_{a}(\omega)-\frac{\Delta_{ab}^{2}(k)}{\omega+\epsilon_{kb}+\Sigma_{b}^{\ast}(\omega)}]^{-1}$
and
$F_{ab}(k,\omega)=G_{aa}(k,\omega)\frac{\Delta_{ab}(k)}{\omega+\epsilon_{kb}+\Sigma_{b}^{\ast}(\omega)}$,
 and introducing the spin operator
$S_{a,\mu}({\bf q})=\frac{1}{2}\sum_{\bf k}c_{a,\mu,\sigma}^{\dag}({\bf k}+{\bf q}){\bf \sigma}_{a,\sigma,\sigma'}^{\mu}c_{a,\mu,\sigma'}({\bf k})$, with $\mu=x,y,z$, the ``bare'' dynamical spin susceptibility reads

\bn
\nonumber
&&\chi_{0,a,b}^{\mu\nu}({\bf q},\omega)=
-\frac{1}{2}{\sigma}_{a,\sigma\sigma'}^{\mu}\cdot{\sigma}_{b,\sigma\sigma'}^{\nu}
\sum_{\bf k,\omega'}[G_{aa}({\bf k}+{\bf q},\omega+\omega') \\ \nonumber
&&G_{bb}({\bf k},\omega')+ F_{ab}(-{\bf k}-{\bf q},-\omega-\omega')F_{ba}({\bf k},\omega')].
\en

  Including the ladder vertex in an infinite summation of ``ladder'' diagrams 
using RPA, the renormalized magnetic susceptibility, $\chi_{a,b}({\bf
q},\omega)=[\chi_{0,a,b}^{-1}(\omega)-J({\bf q})]^{-1}$, where
$\chi_{0,a,b}(\omega)=\sum_{\bf q}\chi_{0,a,b}({\bf q},\omega)$ and 
$J({\bf q})=J_{1}$(cos($q_{x}a$)+cos($q_{y}a$))+$J_{2}$cos($q_{x}a)$cos($q_{y}a$),
with $J_{1}\simeq \frac{t_{ab}^{2}}{U'+J_{H}}$ and $J_{2}\simeq
\frac{t_{ab}'^{2}}{U'+J_{H}}$ being the frustrated superexchange scales in
FePn~\cite{[2]}.
Using $\chi_{0,a,b}(\epsilon)=C\int d\epsilon f(\epsilon)[1-f(\epsilon)]
W(\epsilon)$ in the RPA series,
the NMR relaxation rate, $T_{1}^{-1}=\sum_{\bf q}\frac{\chi''({\bf
q},\omega)}{\omega}|_{\omega\rightarrow 0}$, can be now expressed in terms of 
the {\it full} DMFT propagators.
Here, $W(\epsilon)=\sum_{a,b}[\rho_{aa}(\epsilon)\rho_{bb}(\epsilon)+\rho_{ab}(\epsilon)\rho_{ba}(\epsilon)]$~\cite{[kheli]} and the 
$\rho_{aa}(\epsilon), \rho_{ab}(\epsilon)$ are the LDA+DMFT local spectral functions computed 
earlier~\cite{[18]}.  
Also, $C=2(\frac{2\pi}{\hbar})(\gamma_{e}\gamma_{n}\hbar)^{2}\langle\frac{1}{r^{3}}\rangle$.
 Finally, our restriction to the non-crossing diagrams in the ladder
approximation for $\chi({\bf q},\omega)$ is an approximation.  It is possible
that ``non-crossing'' diagrams need to be included in a full description.
However, for the underdoped cuprates, it has been shown that a renormalized
``RPA'' summation for $\chi({\bf q},\omega)$ with fully renormalized
one-particle $G_{\sigma}(k,\omega),F_{\sigma,-\sigma}(k,\omega)$ gives 
excellent reconciliation
of ARPES and INS data~\cite{[campu]}.  This suggests small vertex corrections: while we
cannot prove why this should be the case, we argue that the good agreement we
find below is an {\it a posteriori} justification for neglecting them in our theory.

The NMR spin-lattice relaxation rate is a measure of the 
{\it local} spin fluctuation rate in both phases.  For an $s$-wave SC, the 
coherence factors give the Hebel-Slichter (HS) enhancement as a peak in 
$T_{1}^{-1}$ below $T_{c}$.  When the ``normal'' state is strongly incoherent 
(large Im$\Sigma(\omega=E_{F})\ne 0$, as in our case), or the SC 
gap has nodes~\cite{[3]}, the HS peak is absent. 
But $T_{1}^{-1}\simeq e^{-\Delta/kT}$ survives for 
$T<<T_{c}$ in an $s$-wave SC, while a power-law fall-off in $T$ characterizes 
an U-SC with gap nodes~\cite{[3],[4]}.  In the normal state above $T_{c}$, we 
set $F_{ab}({\bf k},\omega)=0$.  This suffices for computing the NMR 
$T_{1}^{-1}$.  More work has to be done to compute the INS intensity; we will 
present deatils in a separate work.

\begin{figure}[thb]
\vspace*{-1.5cm}
\hspace*{-3.6cm}
\includegraphics[width=5.7in]{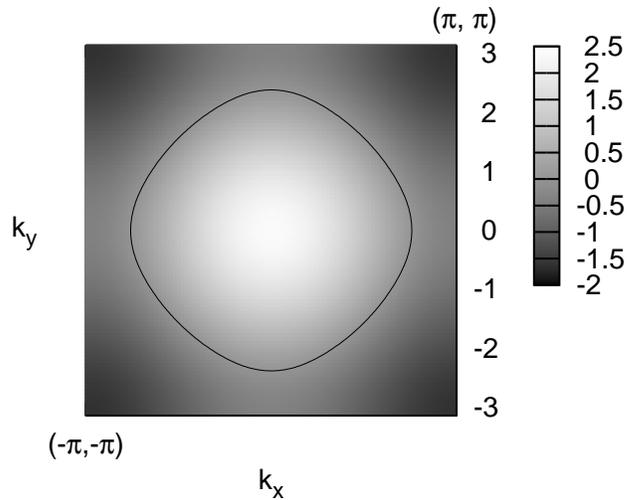}
\vspace*{-1.4cm}
\caption{Proposed gap function for the 1111-Iron Pnictides.  The gap function
has nearest-($\Delta_{1}$, with ex-$s$ symmetry) and next-nearest
($\Delta_{2}$, with $s_{xy}$ symmetry) neighbor components.  With
$\Delta_{2}/\Delta_{1}=0.375$, no in-plane gap nodes arise in the gap
function, in agreement with experiment~\cite{[8]}.}
\label{fig1}
\end{figure}

However, qualitative remarks about what we expect in the INS response are
possible without a full analysis.  The in-plane part, 
$\Delta_{ab}(k)=\Delta_{1}$(cos($k_{x}a$)+cos($k_{y}a$))+$\Delta_{2}$cos($k_{x}a$)cos($k_{y}a$), of our proposed gap function is shown in 
Fig.~\ref{fig1}.  With electron- and hole Fermi 
sheets well separated as in LDA (or LDA+DMFT), no in-plane gap nodes are 
possible, in agreement with a host of measurements~\cite{[1],[8],[17]}.  
Interestingly, this leads to $\Delta({\bf k}+{\bf Q})\Delta({\bf k})<0$ for ${\bf k}$ along $(0,0)-(\pi,\pi)$ and to $\Delta({\bf k}+{\bf Q})\Delta({\bf k})>0$ for
${\bf k}$ near $(\pm\pi/2,0),(0,\pm\pi/2)$.  This implies, following earlier 
work~\cite{[12]}, that INS measurements will show appearance of a 
sharp collective ``spin exciton'' mode in the U-SC state at ${\bf
k}=(\pi,\pi)$, but none 
for ${\bf k}=(\pm\pi/2,0)$.  Of course, incoherent features coming from DMFT 
propagators will introduce damping of this mode, but the qualitative feature
should survive.  Since an out-of-plane cos($k_{z}c$) 
component is induced in the full gap function due to interband 
proximity~\cite{[18]} effect, the INS intensity should also reflect this 
modulation in $q_{z}$.  This last prediction is a consequence of our 
form of the full gap function, and goes beyond previous work~\cite{[12]}.  
Such a resonance, albeit sizably damped, is indeed seen 
in INS work on the 122-FePn~\cite{[10]}.  Moreover, the cos($q_{z}c$) 
form has also been measured by INS on the 122 FePn~\cite{[11]}, but 
remains to be checked in the 1111 family.  Finally, the in-plane angular 
modulation of the gap function is inferred from $\mu$SR work on the 
$Sm$-based FePn~\cite{[9]}.  Thus, rationalization with extant INS 
results readily follows directly from our proposal for the gap function.

Next, we discuss the NMR relaxation rate in the normal and U-SC states, 
making detailed comparison with experimental work.  In the normal, 
incoherent metal state, the reduction in $T_{1}^{-1}$ below $200$~K~\cite{[4],[21]}
indicates opening of a spin gap, as in underdoped cuprates.  While the spin 
gap in cuprates has been identified with short-range magnetic 
correlations in a quasi-$2D$, doped quantum antiferromagnet, its origin in the 
multi-band FePn is not settled.  We emphasize that this behavior is observed in
the same regime where PES~\cite{[8]} and optical data~\cite{[22]} show 
{\it incoherent} charge dynamics, corroborated by a linear-in-$T$ resistivity, 
a $T$-dependent Hall constant, and no Drude peak in optics.  
All these are compelling indicators 
of a strongly correlated metal.  We regard this as a justification for using
LDA+DMFT.  

\begin{figure}[thb]
\includegraphics[width=\columnwidth]{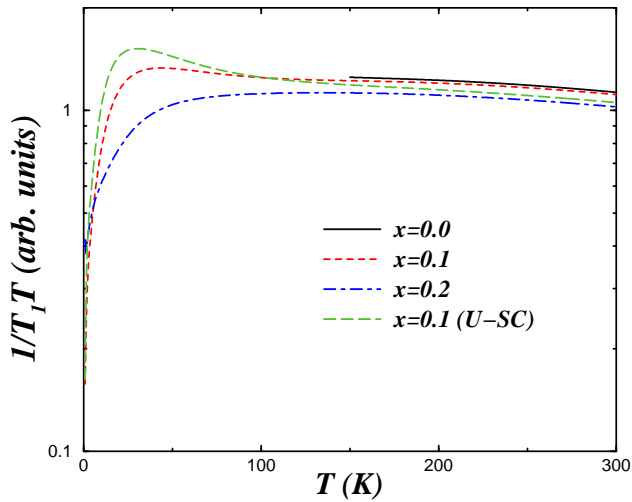}
\caption{(Color online) $T$-dependence of the NMR $(T_{1}T)^{-1}$ over the
full $T$ range for $LaO_{1-x}FeAsF_{x}$, with $x=0$ (black), $x=0.1$ (red, dotted),
$x=0.2$ (blue, dot-dashed) and with inclusion of U-SC for $x=0.1$ (green,
dashed).  Notice how a doping-dependent spin gap around $T^{*}\simeq 150$~K
opens in the doped case ($x=0.1,0.2$), in good agreement with 
experiment~\cite{[4]}.}
\label{fig2}
\end{figure}

In Fig.~\ref{fig2}, we show the NMR $(T_{1}T)^{-1}$ as a function
of electron doping for $LaO_{1-x}FeAsF_{x}$, with $=0.0,0.1,0.2$.  Since we
do not consider the ${\bf q}=(\pi,0)$ SDW phase, the $x=0.0$ curve should only
be trusted above $T=T_{N}=135$~K (shown by the black curve in
Fig.~\ref{fig2}).  With
$x=0.1,0.2$, however, SDW order is destroyed, and U-SC emerges at low $T$.  
In this range of $x$, our results can validly be compared to experiment, which 
we now turn to do.

\begin{figure}[thb]
\includegraphics[width=\columnwidth]{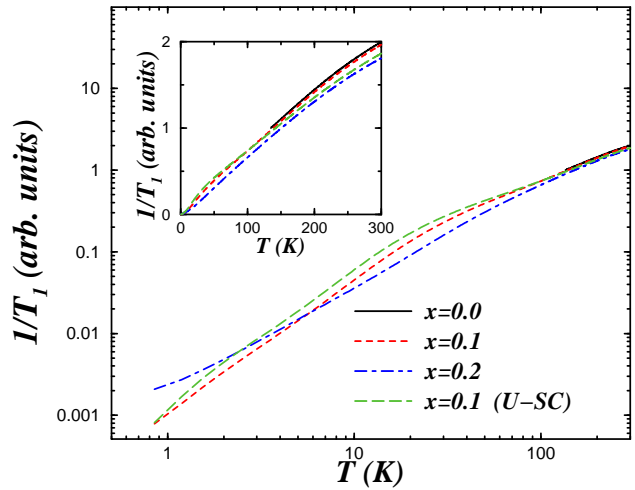}
\caption{(Color online) Low $T$ behavior of the NMR $T_{1}^{-1}$ on a log-log
plot (main panel) and on a normal scale (inset).  Clear power-law behavior
without the Hebel-Slichter coherence peak, in good agreement with
experiment~\cite{[4]}, is seen.   }
\label{fig3}
\end{figure}

Quite remarkably, a direct comparison with published NMR work~\cite{[4]}
reveals good agreement between theory and experiment around $x=0.1$.  
The absence of the $(T_{1}T)^{-1}=const$ regime is striking.
In particular, both experiment and our result show a quasi-linear-in-$T$ (like
 $T^{0.8-0.9}$) increase in $1/T_{1}$ at ``high'' $T>200$~K (see inset of
Fig.~\ref{fig3}).  This resembles 
the high-$T$ precursor of a quantum critical system, and corresponds to the 
``strange metal'' regime in the $T$ vs $x$ phase diagrams for this 
system~\cite{[23]}.  However, as $T$ is lowered, a {\it smooth} drop in 
$(T_{1}T)^{-1}$ around $150$~K marks the onset of the gradual opening up of 
a {\it spin} gap.  Given strong frustration ($J_{2}/J_{1}\simeq 0.7$) in FePn,
 it is tempting to link this spin gap with strong, short-ranged AF
correlations, which are expected to survive the doping induced destruction 
of the SDW~\cite{[24]}.  We
note that the $J_{1}-J_{2}$~\cite{[25]} model has also been used to 
provide a quantitative 
fit of INS results for the undoped 122 FePn, though it is formally valid in the
strictly localized regime.  This is additional evidence for a strong coupling 
picture, since, in the itinerant picture, melting of the SDW should yield a 
paramagnetic Fermi liquid at low $T$ with {\it no} spin gap, at variance with 
observations.  From our results, we estimate a {\it renormalized} spin gap
scale $O(150)$~K in the 1111 FePn.  

In fact, $T_{1}^{-1}\approx tanh(0.42T)$ over almost the whole 
range from low-($T>15$~K) to high $T$.  While this is not particularly 
illuminating, it shows that the ``marginal'' 
form, $\chi_{loc}''(\omega)\simeq -(\omega/T)$, is recovered only at high $T$, 
and is cut off by the spin gap around $200-250$~K.  This bears a peculiar 
resemblance to underdoped cuprates.  However, at very low $T$, a power-law form
$T_{1}^{-1}\simeq T^{1.5-1.6}$, is seen.  This is intriguing, and is fit 
neither by self-consistent renormalization theory~\cite{[26]}, nor by any
known local non-FL exponents~\cite{[27]}.  
It could involve several, frustrated, nearly degenerate spin fluctuation 
channels coming from the multiband nature of FePn, but we are unable to 
quantify this further.

  However, we can still make a few qualitative remarks to get more insight.
In the strongly correlated metal, with a very small ``coherent'' component in
the DMFT spectral functions~\cite{[2],[18]}, ``Mottness'' underpins the low-energy
physics.  More precisely, when one is close to a correlation-driven Mott
insulator, the metallic state has small density of quasi-itinerant carriers
co-existing with effectively local moments~\cite{[2]}.  These latter arise
from integrating out the high energy Hubbard bands in the DMFT spectral
function, as argued by Baskaran, Si {\it et al.} and Wu {\it et al.} Given the
frustrated hoppings characteristic of FePn, the spin degrees of freedom are
qualitatively described by an {\it effective} frustrated $J_{1}-J_{2}$
Heisenberg-type model.  In this model, there is a large window in $T$, between
$T_{SDW}$ and $T_{s}$~\cite{[24]}, where lattice translational symmetry is
spontaneously broken but the spin rotational (SU$(2)$) symmetry is not.  This 
naturally leads to generation of a spin gap, in agreement with observations.
Of course, as LDA+DMFT shows, the actual situation in FePn is somewhat removed
from a strictly localized limit where the $J_{1}-J_{2}$ model would apply.
However, in view of the Mottness, we believe that it still provides a
qualitative understanding of the features derived above in the full DMFT calculation.
 
At $T_{c}$, there is no HS peak, as seen in Fig.~\ref{fig3}: in our work, 
this arises from strong inelastic scattering in the ``normal'' incoherent 
state~\cite{[18],[20]} (notice that $\Sigma_{b}^{*}(\omega)$ enters the DMFT
equation for 
$G_{a}(\omega)$ in the SC state, producing strong damping).  At very low 
$T<<T_{c}$, $T_{1}^{-1}(T)$ shows a power-law-in-$T$ dependence: 
$T_{1}^{-1}(T)\simeq T^{n}$, with $n=2.2-2.5$, qualitatively consistent with 
observations in the 1111 FePn~\cite{[4]}, which show neither a $T^{3}$ nor a 
$T^{5}$ law for $T<<T_{c}$.  The two-step variation of $T_{1}^{-1}$ below
$T_{c}$ is also reproduced theoretically.  The first ``step'' from
$T_{c}>T>T_{c}/4$ is dominantly governed by the larger gap component, while
the lower-$T$ variation comes from the smaller gap component, as expected from
an in-plane anisotropic gap, while the power-law variation is ascribed to
out-of-plane line nodes in such a gap.  It is still possible that disorder 
effects (which must be treated in the unitary limit~\cite{[1]}) will lift the 
out-of-plane gap 
nodes, as discussed by Maier {\it et al.}~\cite{[12]} and give
$T_{1}^{-1}\simeq T^{3}$ behavior~\cite{[6]}; 
this remains to be checked.  In our theory, the power law behavior arises from the 
out-of-plane line nodes, induced in the gap by an interband proximity effect.
Since it does not {\it require} disorder effects, our conclusion should be more 
``universal''~\cite{[16]}.  Thus, our results show how good agreement with the
NMR data is derived in the whole $T$ range in terms of our theoretical picture
of an U-SC with proximity induced line nodes, arising from an incoherent
normal state at $T_{c}$.  

In conclusion, we have studied the magnetic fluctuations in FePn, based on a 
novel theoretical proposal for the symmetry of the SC gap function.  In a picture
where U-SC with out-of-plane gap nodes arises from an incoherent, strongly
correlated normal state, we have shown how the $T$-dependence of the NMR 
relaxation rate can be nicely understood over the whole $T$ range, from the 
lowest- to ``high'' $T$.  Moreover, we have argued how the specific form of 
the gap function allows for concrete predictions concerning the observation of 
the collective resonant peak in INS measurements.  Our study provides further
support for the strongly correlated nature of FePn above $T_{c}$, and puts our
theoretical proposal of an U-SC with out-of-plane gap nodes on a firmer 
footing.


\end{document}